# HIGH DIELECTRIC SHEET TO REDUCE ELECTRIC FIELDS AND FLATTEN MAGNETIC FIELDS IN SELF-DECOUPLED RADIOFREQUENCY COILS FOR MR IMAGING

Aditya Bhosale[1], Xiaoliang Zhang[1,2]

[1]Department of Biomedical Engineering, [2]Department of Electrical Engineering, State University of New York at Buffalo, NY 14260, USA

*Abstract*— **High impedance RF coils, such as self-decoupled coils, reduce the electromagnetic coupling between the coil elements and eliminates the use of complex decoupling technologies. Although the high impedance design promises excellent decoupling between the coil elements, it also results in high electric fields across the RF coil, leading to potential safety problems during imaging. It also causes B1 field asymmetry, ultimately leading to difficulties in imaging quantification. In this study, we propose and investigate using a high dielectric sheet to reduce the electric fields across the coil and achieve excellent electromagnetic decoupling among the coil elements, thereby ensuring safer MRI at ultra-high fields and maintaining high imaging performance.**

## I. INTRODUCTION

In magnetic resonance imaging, arrays of radiofrequency coils are commonly used to achieve high signal-to-noise ratios and versatile volume coverage, speed up scans with parallel reception, and reduce field nonuniformity with the parallel transmission. On the other hand, traditional coil arrays require sophisticated decoupling technologies to minimize electromagnetic coupling between coil elements, which would otherwise amplify noise and reduce transmitted strength [1-14]. Electromagnetic coupling between RF array elements has posed a daunting challenge in designing multichannel coil arrays, particularly transceiver arrays at high and ultrahigh fields [15-27].

The recently proposed self-decoupled coil is a unique alternative to standard coils. The self-decoupled coils have a simple structure and deliberate electrical impedance redistribution along the coil loop's length. Self-decoupled coils have excellent decoupling efficiency despite their coil-orientation dependence. When used to construct an array, they are resistant to coil separation, making them appealing for size adjustable and versatile coil arrays. Easy rectangular or circular loop coils with evenly spaced capacitors connected along their length are known as conventional coils.

On the other hand, the self-decoupled coil has the same structure as a traditional coil. A comparatively high impedance is located opposite the coil's feed port, causing the coil to act like a loop and a folded dipole superimposed on each other [28-30]. Due to the small value capacitance (~0.4pF) in the self-decoupled coil, a high electric field will be generated in the area around the capacitor. This may increase the Specific Absorption Rate (SAR), resulting in an elevated tissue-heating hazard during MR imaging examinations. The high electric fields caused by the small value capacitance can be mitigated using multiple series capacitors with relatively large capacitance values. This work proposes and investigates a method using high permittivity dielectric materials to suppress the electric field around the capacitor. Due to its unique property, the radio frequency magnetic field (B1 field) is expected not to be negatively impacted.

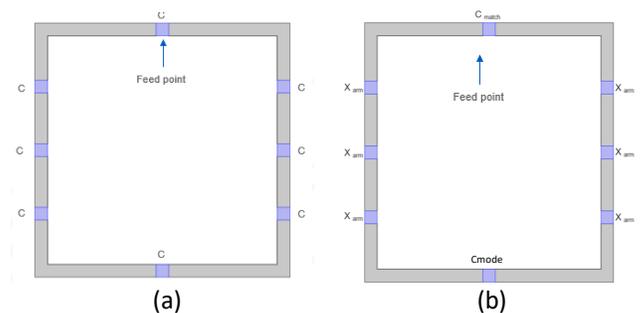

(a)       (b)

Fig. 1 (a) Conventional coil with eight capacitors (each 8 pF) spaced evenly along its length (b) Self-decoupled coil high impedance ($C_{mode}$:0.44pF) placed opposite the coil's feed port.

## II. METHODS

We modeled a conventional coil, self-decoupled coil, and self-decoupled coil with a dielectric sheet. The convention coil has eight 8pF capacitors equally distributed along its length. The self-decoupled coil comprises six $X_{arm}$ impedances, one $C_{mode}$ capacitor opposite the feed port, and one $C_{match}$ capacitor to balance the self-decoupled coil's impedance. The magnetic and electric field distribution generated through the self-decoupled coil was simulated and evaluated using a $10\times10$ cm$^2$ self-decoupled coil. The safety concern associated with the self-decoupled coil is the higher electric fields produced across the $C_{mode}$ capacitor of the self-decoupled coil. One method to reduce the higher electric fields produced across the coil is to break up the $C_{mode}$ capacitor into five capacitors of 2.9pF. In this study, we propose another method that uses a high dielectric sheet to reduce the higher electric fields across the coil. We designed a $10\times10$ cm$^2$ self-decoupled coil and mounted a 1mm thick piece of the dielectric sheet over the Cmode capacitor to see how the high dielectric sheet affected the electric, magnetic, and decoupling performance of the coil [13-15, 31, 32].

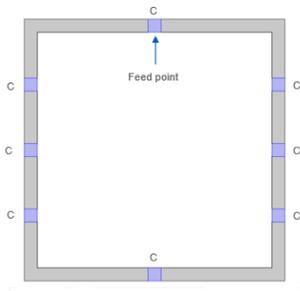

Fig. 2 Conventional coil with eight capacitors equally distributed along its length.

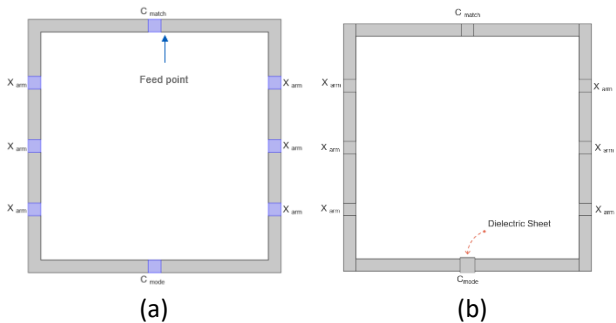

Fig. 3 (a) Self-decoupled coil (b) Dielectric sheet over $C_{mode}$ capacitor of Self-decoupled coil.

### C. High Dielectric Sheet

We placed a high dielectric sheet with a thickness of 1mm above the $C_{mode}$ capacitor. The dimensions of the dielectric sheet were $1\times1$ cm$^2$. The high dielectric sheet's relative permittivity was varied to test the array system's performance at different values. Figure 3 (b) shows the dielectric sheet above the self-decoupled coil.

### D. Simulations

For a $10\times10$ cm$^2$ self-decoupled coil without a dielectric layer, each $X_{arm}$ impedance is 5.6nH & the value of the $C_{mode}$

capacitor is 0.44pF. The $C_{match}$ capacitor is solely used to match the coil's impedance at 50ohm. We ran simulations with a small piece of the dielectric sheet covering the $C_{mode}$ capacitor on the self-decoupled coil. The dielectric sheet's relative permittivity varied from 50 to 400. The resonant frequency of the self-decoupled coil was matched at 300MHz/50 ohm by tuning the $X_{arm}$ impedances and the $C_{match}$ capacitor to compensate for the dielectric sheet load. The self-decoupled coil's input power was kept constant at 1 Watt. The effect of the dielectric sheet on the electric fields and magnetic fields across the self-decoupled coil was observed. The decoupling efficiency was also observed using electromagnetic simulations.

### E. Data Analysis

We exported the B field maps, current density plots, and S-parameters directly using COMSOL Multiphysics software. The magnetic field distribution was displayed using the following expression:

emw. normB

The surface current density plots were constructed using the following expression:

emw.Jsurf

The Electric field plots were constructed using the following expression:

emw.normE

## III. RESULTS

### A. Electric Field Distribution

The electric field of the conventional coil is $1.8\times10^3$ V/m, while the electric field across the $C_{mode}$ capacitor on the self-decoupled coil is $7.4\times10^3$ V/m, which is almost four times higher.

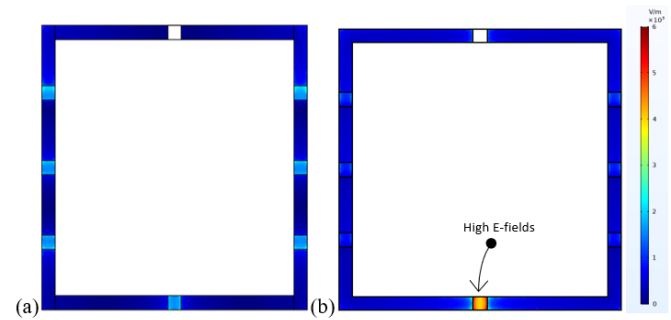

Fig. 4 Electric field distribution (a) Conventional coil (b) Self-decoupled coil

Our results show a steady decrease in the electric fields across the $C_{mode}$ capacitor on the self-decoupled coil with a higher value of relative permittivity of the dielectric sheet. The electric field strengths across the $C_{mode}$ capacitor for all the tested values of the dielectric sheet's relative permittivity are listed in the table below[33-37].

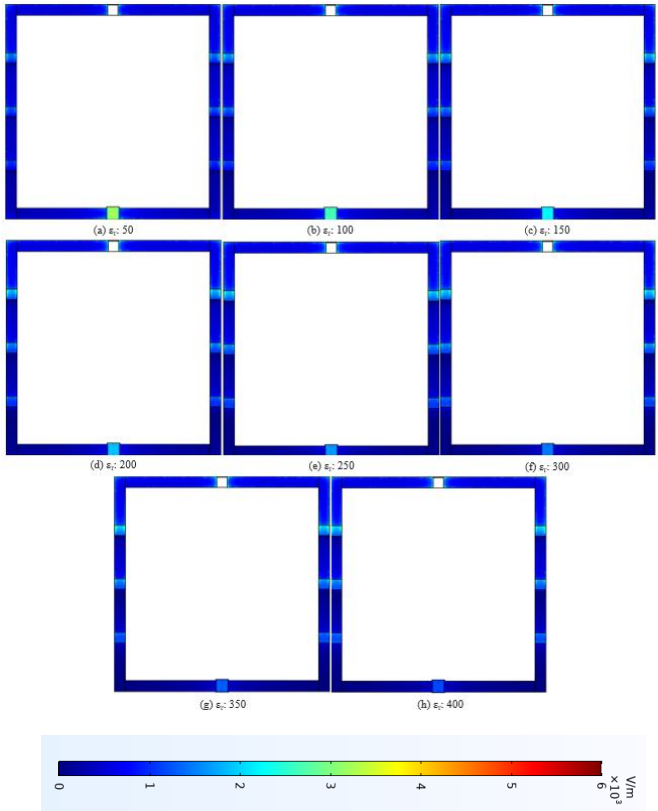

Fig. 5 Electric field distribution of a self-decoupled coil with dielectric sheet over $C_{mode}$ capacitor (a) $\varepsilon_r$: 50 (b) $\varepsilon_r$: 100 (c) $\varepsilon_r$: 150 (d) $\varepsilon_r$: 200 (e) $\varepsilon_r$: 250 (f) $\varepsilon_r$: 300 (g) $\varepsilon_r$: 350 (h) $\varepsilon_r$: 400

| Coil Type | Electric fields across $C_{mode}$ capacitor |
|---|---|
| Self-decoupled coil without a dielectric sheet | $7.4 \times 10^3$ V m$^{-1}$ |
| Self-decoupled coil with a dielectric sheet ($\varepsilon_r$ = 50) | $3.2 \times 10^3$ V m$^{-1}$ |
| Self-decoupled coil with a dielectric sheet ($\varepsilon_r$ = 100) | $2.5 \times 10^3$ V m$^{-1}$ |
| Self-decoupled coil with a dielectric sheet ($\varepsilon_r$ = 150) | $2.1 \times 10^3$ V m$^{-1}$ |
| Self-decoupled coil with a dielectric sheet ($\varepsilon_r$ = 200) | $2 \times 10^3$ V m$^{-1}$ |
| Self-decoupled coil with a dielectric sheet ($\varepsilon_r$ = 250) | $1.8 \times 10^3$ V m$^{-1}$ |
| Self-decoupled coil with a dielectric sheet ($\varepsilon_r$ = 300) | $1.5 \times 10^3$ V m$^{-1}$ |
| Self-decoupled coil with a dielectric sheet ($\varepsilon_r$ = 350) | $1.2 \times 10^3$ V m$^{-1}$ |
| Self-decoupled coil with a dielectric sheet ($\varepsilon_r$ = 400) | $1 \times 10^3$ V m$^{-1}$ |

Table. 1 Effect of the dielectric sheet on electric field strength across $C_{mode}$ capacitor on the self-decoupled coil

## B. Magnetic Field Distribution

The magnetic field distributions of the conventional coil, self-decoupled coil, and self-decoupled coil with dielectric sheet are shown using the linear scale in T. The expression used is as follows: (emw.normB).

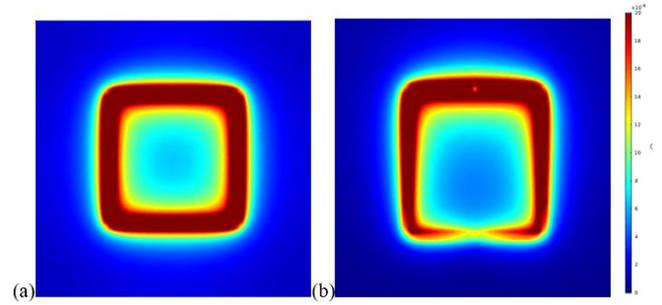

Fig. 6 Magnetic field distribution (a) Conventional coil (b) Self-decoupled coil

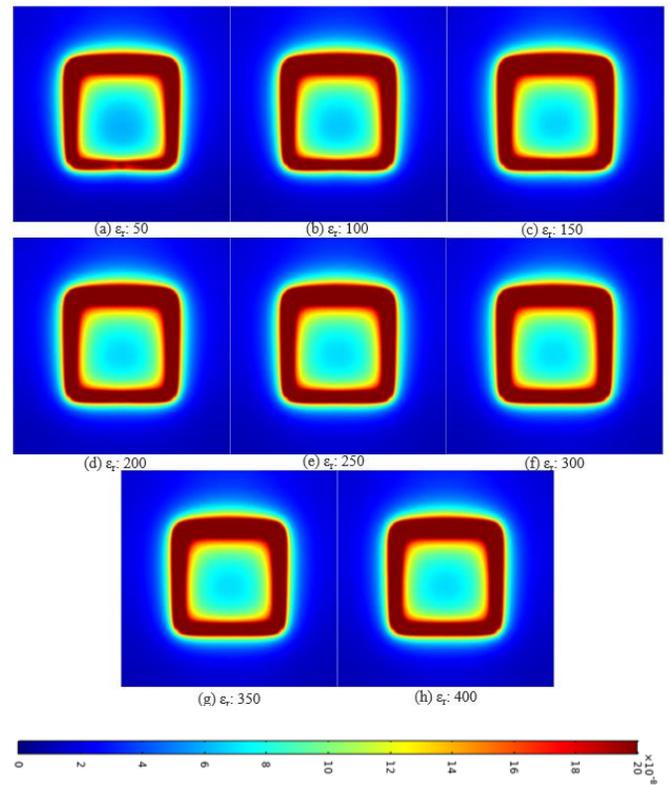

Fig. 7 Magnetic field distribution of a self-decoupled coil with dielectric sheet over $C_{mode}$ capacitor (a) $\varepsilon_r$: 50 (b) $\varepsilon_r$: 100 (c) $\varepsilon_r$: 150 (d) $\varepsilon_r$: 200 (e) $\varepsilon_r$: 250 (f) $\varepsilon_r$: 300 (g) $\varepsilon_r$: 350 (h) $\varepsilon_r$: 400

Our findings suggest that as the relative permittivity of the dielectric layer increases, the electric fields decrease, and the magnetic field distribution becomes more uniform and balanced.

## C. Decoupling Performance

The current distributions of the self-decoupled coil with the dielectric sheet on the $C_{mode}$ capacitor are shown using the linear scale in A/m. The expression used is as follows: (emw.Jsurf). The distance between two self-decoupled coils is 1 cm. One coil is excited, while the other is kept as a passive element. The results suggest that using a high dielectric sheet does not hamper the decoupling efficiency of the self-decoupled coil, and the current induced in the passive element is negligible. Hence, our proposed method promises excellent decoupling efficiency.

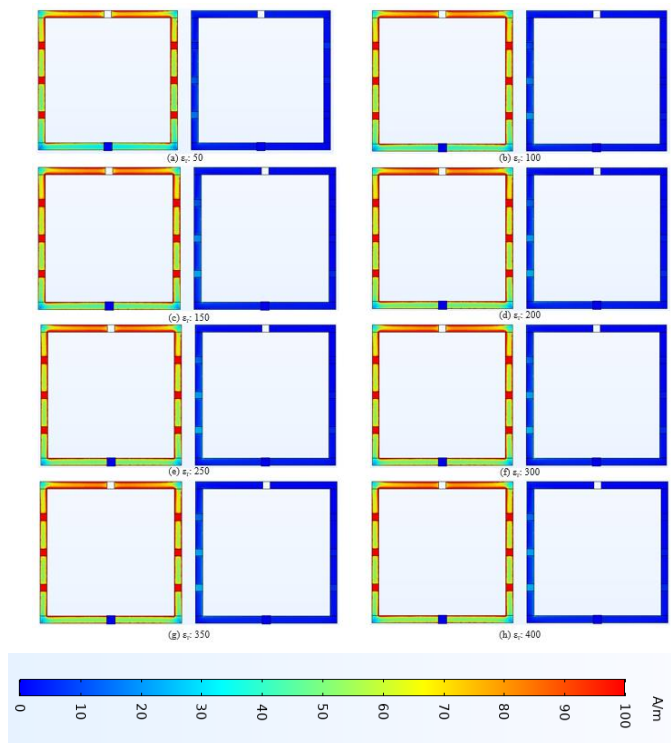

Fig. 8 Simulated current distributions of the self-decoupled coils with a dielectric sheet cover the Cmode capacitor where one coil is excited. The other coil is kept as a passive element. (a) $\varepsilon_r$: 50 (b) $\varepsilon_r$: 100 (c) $\varepsilon_r$: 150 (d) $\varepsilon_r$: 200 (e) $\varepsilon_r$: 250 (f) $\varepsilon_r$: 300 (g) $\varepsilon_r$: 350 (h) $\varepsilon_r$: 400

## IV. CONCLUSION AND DISCUSSION

The self-decoupled coil's high dielectric sheet significantly decreased electric fields, accounting for safer MRI at ultra-high fields. It also improved the magnetic field distribution while ensuring decoupling efficiency. In the future, Bench tests may be carried out to test our proposed method's success in a realistic setting.

## Acknowledgment

This work is supported in part by the grant from NIH (U01 EB023829) and SUNY Empire Innovation Professorship Award.